\documentstyle[prb,preprint,aps]{revtex}    

\begin{document}
\draft
\newcommand{\beq}{\begin{equation}}
\newcommand{\eeq}{\end{equation}}
\newcommand{\beqa}{\begin{eqnarray}}
\newcommand{\eeqa}{\end{eqnarray}} 

\title{Absence of charge backscattering\\
in the nonequilibrium current of\\
normal-superconductor structures}

\author{J. S\'anchez Ca\~nizares and F. Sols}
\bigskip
\address{
Departamento de F\'{\i}sica Te\'{o}rica de la Materia Condensada, C-V, and\\
Instituto Universitario de Ciencia de Materiales ``Nicol\'as Cabrera''\\
Universidad Aut\'onoma de Madrid, E-28049 Madrid, Spain}

\maketitle
\begin{abstract}

We study the nonequilibrium transport properties of a
normal-superconductor-normal structure, focussing
on the effect of adding an impurity in the superconducting 
region. Current conservation requires the superfluid
velocity to be nonzero, causing a
distortion of the quasiparticle dispersion relation within
the superconductor. For weakly reflecting
interfaces we find a 
regime of intermediate voltages in which Andreev transmission
is the only permitted mechanism for quasiparticles to enter
the superconductor. Impurities in the superconductor 
can only cause Andreev reflection of these quasiparticles
and thus cannot degrade the current.
At higher voltages, a state of gapless
superconductivity develops which
is sensitive to the presence of impurities.

\end{abstract}
\vspace{.5cm}


\vspace{.5cm}
\narrowtext

During the last few years, 
considerable progress has been made
in the understanding of
transport in
mesoscopic structures with
superconducting elements.\cite{karls} The 
possibility of phase-coherent Andreev 
reflection \cite{andreev} 
provides a rich variety of situations with novel
transport anomalies.\cite{review}
It has been pointed out \cite{sanc95,mart95} that
a limitation of conventional descriptions of
Andreev reflection is that, by assuming
a uniform phase in the superconducting region, they
implicitly contain a violation of current conservation.
While this shortcoming may be unimportant
in the regime of low current densities, it certainly
cannot be overlooked in situations where the superconductor
supports current densities that are comparable to 
its equilibrium critical value. This may easily be the case
in transmissive structures with applied
voltages $V \sim \Delta_0/e$, where $\Delta_0$ is the
zero current superconducting gap. It has been
shown \cite{sanc95,mart95} that
the introduction of a nonuniform superconducting phase
to ensure self-consistency (and, with it, charge conservation
\cite{furu91,sols94,bagw94}) may have major consequences
in the current-voltage characteristics of NS
and NSN structures. Structures with weakly reflecting interfaces
present a regime of intermediate voltages in which Andreev
transmission (whereby an electron is transmitted as
a quasihole, or viceversa) is the only allowed mechanism for
quasiparticles to enter the S region. Here we
wish to analyse this transport regime (hereafter referred to
as the AT regime), which is
characterized by the exclusive presence of Andreev transmitted
quasiparticles. In particular,
we report on the existence in the AT regime
of a curious effect by which impurities are unable
to degrade the quasiparticle electric current
in the superconducting region. The origin of this effect
is the strong distortion that the quasiparticle
dispersion relation suffers in the presence of a 
condensate with a large superfluid velocity $v_s$, as is
shown schematically in the inset of Fig. 1a. 
The energy thresholds for quasiparticle propagation
become $\Delta_{\pm}=\Delta_0 \pm \hbar v_F q$ for
$k \simeq \pm k_F$, with $k_F=mv_F/\hbar$ the Fermi
momentum and $q=mv_s/\hbar$. 
In a NSN structure,
at zero temperature, and if both NS interfaces
are identical, electrons(holes) come from the left(right) N lead
with energies $0<\varepsilon<eV/2$, and
the AT regime is characterized by the condition
$0<\Delta_-<eV/2<\Delta_+$,
where $V$ is the total voltage
difference between the two N leads.
At these intermediate voltages, the only scattering mechanisms 
available for
electrons coming from the left N lead 
are normal reflection (NR), Andreev reflection (AR),
and Andreev tranmission (AT).
A similar consideration applies for holes coming
from the right N lead.
The current in S is then carried 
by the condensate and by quasiparticles
lying exclusively in the low energy branch of the 
asymmetric $\varepsilon(k)$ curve. 
One such quasiparticle (i.e., a
right-moving quasihole) may undergo elastic scattering by impurities, 
with the peculiarity that, at energies $\Delta_- <\varepsilon <\Delta_+$,
the only available scattering channels are either
normal transmission (NT) as a right-moving quasihole or 
AR as a left-moving
quasielectron.\cite{comm5} It is clear that in both cases 
the electric current is left intact (up to corrections of
order $\Delta_0/E_F$). We conclude that, 
in the AT regime of intermediate voltages, 
the quasiparticle
component of the electric current is essentially insensitive to
the presence of impurities. In this article we
present numerical results for a one-dimensional model
that support this qualitative
prediction. We also argue that the main conclusion
concerning the absence of quasiparticle current
degradation must apply under more realistic 
circumstances.

We solve
the Bogoliubov -- de Gennes equations \cite{dege66}
for a chosen one-dimensional structure with the requirement that
the self-consistency condition
\beq
\Delta = g \sum_{\alpha} u_{\alpha}  
v_{\alpha} ^* (1 - 2 f_{\alpha} ),
\eeq
is satisfied ($(u_{\alpha},v_{\alpha})$ and $f_{\alpha}$
are the spinor wave function and the occupation probability
for quasiparticle $\alpha$).
A number of assumptions are similar to those
of Ref. \cite{sanc95}. The gap function is taken to 
be of the form
$\Delta(x)=|\Delta|e^{2iqx}$,
where the parameters $|\Delta|$ and $q$ 
are determined self-consistently from Eq. (1) and 
from the conservation of current. 
Details of the numerical calculation will be given
in a forthcoming publication.\cite{sanc95a}
The gap amplitude 
$|\Delta|$ is assumed to be uniform within the superconductor.
This is expected to be a reasonable approximation for structures
longer than a few times the zero-temperature
coherence length $\xi_0=\hbar v_F/\Delta_0$, so
that $|\Delta(x)|$ can relax to its bulk saturation
value. \cite{mart95}
We introduce one-electron delta function barriers of
strength $Z$ at the NS interfaces \cite{sanc95,mart95,blon82}
and one barrier of strength $Z_I$ to model the impurity within
the superconductor.
\cite{bagw93}
We assume incoherent (albeit elastic)
multiple scattering
by the impurity and at the interfaces. 
This requires the existence of
some type of phase breaking mechanism involving
negligible energy loss for the quasiparticles.
This set of simplifying assumptions allows us
to focus on the 
physics introduced by the combined requirements
of current conservation and impurity scattering 
at high current densities.
The main physical effects come from the splitting
of thresholds (from $\Delta_0$ to $\Delta_{\pm}$
as $v_s$ becomes nonzero), which gives rise to
a variety of transport regimes. 

We concentrate on
the case of two moderately transmissive interfaces
($Z\!=\!0.5$), since it is in this case
where the different transport regimes can be most
clearly identified. \cite{comm1} 
The case of $Z\!=\!0.5$ and $Z_I=0$ was already
studied in Ref. \cite{sanc95} and we only review 
the main facts here.
The AT channel opens for $v\!\equiv eV/\Delta_0\!\simeq\!
1.3$, as can be seen from the jump in the total
current (see Fig. 1a) and from the onset of a negative
quasiparticle current (Fig. 1b). One has $I_{qp} < 0$ because,
in the available energy range 
($\Delta_- < \varepsilon < eV/2 < \Delta_+$),
the only quasiparticle
channels are either right-moving holes or left-moving
electrons, 
all contributing negatively to the current (here, $e>0$).
The change in the total current is positive 
because of the large 
increase in the
condensate current (the sudden jump in $I_{qp}$ at 
$v\!\simeq\!1.45$
is inconsequential for the total current \cite{sanc95}).
The transition to the AT regime is accompanied
by an enhancement
of the superfluid velocity
(see Fig. 2a) and a depression of the gap (Fig. 2b). 
The regime of gapless superconductivity
(GS) sets in
for $v\!\simeq\!
1.8$, as can be guessed from the smooth increase
in $I_{qp}$. 

Let us now study the effect of adding
an impurity ($Z_I \neq 0$)
in the superconducting region. 
Although, for numerical simplicity, we present
a one-dimensional calculation, the type of system we
have in mind 
is a realistic three-dimensional superconductor
with impurities or barriers that may cause scattering 
of quasiparticles.
Thus, we neglect here any possible role of the
impurity as a phase-slip center.
We rather view the impurity
as a mere source of incoherent quasiparticle scattering that 
affects the condensate only indirectly through the self-consistency
condition (1). 

At low voltages,
charge transport is dominated by Andreev reflection.
Current in the superconductor is 
entirely carried by the condensate and thus is not
affected by impurity scattering.
As in case of the clean structure, the AT channel opens at
$v\!\simeq\!1.3$, causing a marked increase in the total
current. Due to the presence of quasiparticles one
might naively expect the $I-V$ characteristics to be sensitive
to the presence of an impurity in the AT regime. However, the peculiar
form of the quasiparticle dispersion relation 
makes this presence ineffective. As already
pointed out, in this regime
only the low-lying states of the left branch are populated.
Scattering by the impurity is unable to degrade the current
carried by these quasiparticles, because only NT and AR are
pemitted and both mechanims
leave the sign of the electric current unaltered.
Due to flux conservation and to the additional symmetry 
stemming from the two NS interfaces being
identical, one finds that
all the magnitudes depending on the quasiparticle
population in the superconductor 
become impurity independent. This fact can be clearly appreciated 
in all the figures if one inspects the AT 
range $1.3 \alt v \alt 1.8$.

The situation changes dramatically when $\Delta_-$ becomes
negative and the state of gaples superconductivity is established.
\cite{sanc95}
As the characteristic
unconventional branch emerges at $k\simeq k_F$,
the four
quasiparticle channels become available at low
enough energies. At this point,
normal reflection by the impurity becomes possible and,
with it, a strong degradation of the current.
This fact is clearly observed in the $v \agt 1.8$ sector
of Fig. 1a. The dependence of the transport properties
in the GS state on
the impurity scattering strength $Z_I$
constrasts
markedly with the insensitivity found in the low voltage regime:
the stronger $Z_I$, the larger the decrease
in the current.
The decrease in the current 
is caused by both
the appearance of NT into the
unconventional gapless branch as the
dominant current-carrying channel at the NS 
interfaces and the possibility of impurity-induced
NR within the S region. 

We also observe in Fig. 1b 
a sharp rise in $I_{qp}$ that
contrasts with the smooth behavior of the 
clean system.
These abrupt features in the current-voltage characteristics
are also related to 
the jumps appearing
in the self-consistently calculated values of the 
superfluid velocity and order parameter (see Fig. 2).
The detailed nature of the sharp transition to the GS
state seems to be very sensitive to 
the impurity strength.
The reason why the size of the gap
discontinuity increases with $Z_I$
(see Fig. 2b) is similar to that which
explains the behavior of a clean NSN structure
with high $Z$. \cite{sanc95}
The population of 
quasiparticles placed in the
emerging unconventional branch tends to reinforce 
the gap. In the presence of strongly reflecting 
barriers, however,
the typical residence time of quasiparticles in the
S region tends to increase and the resulting
directional randomization tends to reduce $|\Delta|$. 

One may wonder whether the
physical effects discussed here would survive in a 
realistic structure with many transverse channels.
Fortunately, the answer is yes, as can be
inferred from the following argument 
(a numerical confirmation will be presented in
Ref. \cite{sanc95a}):
the main
complication caused by the presence of many transverse
channels is that, for a given voltage, different
channels may be in different transport regimes. This
occurs because the shifted energy thresholds, 
$\Delta^{(n)}_{\pm}=\Delta_0 \pm \hbar q v_F^{(n)}$,
depend on the index channel $n$ ($v_F^{(n)}$ is the velocity
available for the longitudinal propagation of Fermi electrons
in channel $n$). 
Just above the low voltage region
dominated by AR,
there must be a range of voltages in which 
scattering of quasiparticles into some channels within S
(those with the highest values of $v_F^{(n)}$) is possible,
but only by Andreev transmission.
Even if impurities induce mixing among those modes,
the basic effect of the absence of current degradation should
remain, since the presence of many transverse channels does not
change the fact that, in that voltage range, only right-moving
quasiholes or left-moving quasielectrons are available. Therefore,
one expects to observe the same kind of impurity independence
that is found for the single channel case. 

Another potential problem is posed by the 
possibility that spatial variations
in the effective quasiparticle chemical 
potential may induce a time-dependent
response of the condensate. 
However, a gradient in the condensate chemical potential
causes an increasing spatial variation of the phase that must
be compensated by an unwinding mechanism. This requires
the presence of phase-slip centers or a
temperature very close to $T_c$, \cite{lang67}
and both factors are ruled out here. The situation is
rather one in which a charge imbalance develops between the
condensate uniform chemical potential and the nonequilibrium
population of quasiparticles. \cite{clar72,revi86} 
The charge imbalance relaxation time,
$\tau_{\varepsilon}$,
must be much longer than the average residence time 
$\tau_r \agt L/v_F$.
On the other hand,
$L \agt \xi_0$ 
is required for $|\Delta(x)|$ to reach its asymptotic 
value.\cite{mart95} Thus, we require
$\tau_{\varepsilon} \gg \hbar/\Delta_0$, which is physically realizable
at sufficiently low temperature.\cite{likh79}

Finally, an important consequence of nonequilibrium is 
the absent of efficient heating.
As in other experimental \cite{rodr94,poza95} 
and theoretical \cite{sanc95,mart95,gray}
works, we find that  voltage differences can coexist with
superconductivity, even when $eV$ is several times $\Delta_0$.
This is due to the lack of complete thermalization of quasiparticles in
the S region.

From the analysis presented here
we may envisage the following experimental
scenario. A finite S segment is inserted in
a N lead of the same or greater width. Contacts
must be good to ensure that, at low temperatures,
$q$ becomes comparable to $\Delta_0/\hbar v_F$
when $V$ is comparable to $\Delta_0/e$. As in
other mesoscopic transport contexts, voltage sources yielding
thermal populations of incoming quasiparticles with a
well-defined chemical potential can be obtained by inserting
the NSN structure in a wide circuit with low impedance.

At low voltages, AR dominates, the current is fully carried
by the condensate and transport is
insensitive to the impurity strength. 
As the voltage increases, a peak in the differential
conductance signals the onset of quasiparticle Andreev tranmission.
The AT regime is also insensitive to the presence
of the impurity. 
At higher voltages, trasnsverse modes with
a high longitudinal Fermi velocity become gapless.
The onset of GS 
should be clearly
identified through a reduction in the current of
a magnitude that increases with $Z_I$. This behavior
contrasts with that expected for structures in which 
the superfluid velocity remains negligible.
That could be the case, for example, in
a NSN structure in which the S segment is wider than the N leads.
For such devices one expects that AT and conventional NT enter into
action simultaneously at a voltage 
$V=2\Delta_0(T)/e$, regardless of the value of $Z$. With all
quasiparticle channels available, one expects the behavior
at voltages just above the first peak in $dI/dV$ to be sensitive 
to the impurity strength $Z_I$.

In conclusion, in this paper we have studied the
transport of current in an
incoherent NSN structure
with an added scattering source in  
the superconducting segment. Nontrivial effects
appear when the current
is large enough to make a self-consistent
calculation necessary. The presence of a nonzero
superfluid velocity creates a distortion in the quasiparticle
dispersion relation that is responsible for the existence
of several transport regimes. These regimes respond to the
presence of impurities in different ways.
In the regime where only Andreev transmitted quasiparticles
are permitted in the superconductor,
all the relevant magnitudes in the
problem are essentially impurity-independent. This happens because the 
impurity (or the set of impurities) is unable to normally reflect a 
quasiparticle
and thus cannot degrade the electric current.
At high voltages, when gapless superconductivity is reached,
normal transmission at the interface and normal reflection
by the impurity become possible, leading to a notable 
decrease in the total current that grows with the
effective scattering strength.

We wish to thank C.J. Lambert, A. Martin,
J.G. Rodrigo, and M. Poza
for valuable discussions. This project
has been supported by Direcci\'on General
de Investigaci\'on Cient\'{\i}fica
y T\'ecnica, Project no. PB93-1248, and by
the Human and Capital Mobility Programme of the EC.
J.S.C. acknowledges support from Ministerio de
Educaci\'on y Ciencia through a FPI fellowship.

\begin{figure}
\caption{
(a) Self-consistent I--V characteristics 
of a NSISN structure
for $Z=0.5$ and six different values of $Z_I$. 
The solid
line corresponds to the case of 
a clean ($Z_I=0$) NSN structure. Inset: schematic
representation of the quasiparticle relation
$\varepsilon(k)$ distorted by a nonzero $v_s$.
Part (b) shows the 
quasiparticle component of the current.
}
\end{figure} 

\begin{figure}
\caption{
Same as Fig. 1, for the 
superfluid velocity (a)
and the amplitude of the
gap (b) in the
superconducting region. $v_s$
is given in units of the depairing velocity
$v_d\equiv \Delta_0/\hbar k_F$.
}
\end{figure}


\begin{references}
\bibitem{comm1} For large $Z$, 
it was found in Ref. 4 that, as the voltage increases,
the system bypasses the AT regime, jumping
directly from the AR dominated region to the
normal resembling sector of gapless 
superconductivity. For vanishing Z, AT is inhibited, since
it is proportional to $r^{*}t$, where $r(t)$ if the one-electron
reflection(transmission) coefficient.
\end{references}
\end{document}